\pdfoutput=1
\documentclass[12pt,preprint]{aastex}

\slugcomment{}

\shorttitle{STELLAR VARIABILITY IN THE CSTAR FIELD}
\shortauthors{WANG ET AL.}

\begin{document}

%% LaTeX will automatically break titles if they run longer than
%% one line. However, you may use \\ to force a line break if
%% you desire.

%\title{Collapsed Cores in Globular Clusters, \\
%    Gauge-Boson Couplings, and AAS\TeX\ Examples}

\title{PHOTOMETRIC VARIABILITY IN THE CSTAR FIELD: RESULTS FROM THE 2008 DATA SET}

%% Use \author, \affil, and the \and command to format
%% author and affiliation information.
%% Note that \email has replaced the old \authoremail command
%% from AASTeX v4.0. You can use \email to mark an email address
%% anywhere in the paper, not just in the front matter.
%% As in the title, you can use \\ to force line breaks.

\author{Songhu Wang\altaffilmark{1}, Hui Zhang\altaffilmark{1}, Xu Zhou\altaffilmark{2}, Ji-Lin Zhou\altaffilmark{1}, Jian-Ning Fu\altaffilmark{3}, Ming Yang\altaffilmark{1}, Huigen Liu\altaffilmark{1}, Jiwei Xie\altaffilmark{1}, Lifan Wang\altaffilmark{4}, Lingzhi Wang\altaffilmark{2}, R. A. Wittenmyer\altaffilmark{5}, M. C. B. Ashley\altaffilmark{5}, Long-Long Feng\altaffilmark{4}, Xuefei Gong\altaffilmark{6}, J. S. Lawrence\altaffilmark{5,7}, Qiang Liu\altaffilmark{2}, D. M. Luong-Van\altaffilmark{5}, Jun Ma\altaffilmark{2}, Xiyan Peng\altaffilmark{2}, J. W. V. Storey\altaffilmark{5}, Zhenyu Wu\altaffilmark{2}, Jun Yan\altaffilmark{2}, Huigen Yang\altaffilmark{8}, Ji Yang\altaffilmark{4}, Xiangyan Yuan\altaffilmark{6}, Tianmeng Zhang\altaffilmark{2}, Xiaojia Zhang\altaffilmark{9}, Zhenxi Zhu\altaffilmark{4}, AND Hu Zou\altaffilmark{2}}
\email{zhouxu@bao.ac.cn, zhoujl@nju.edu.cn}

\altaffiltext{1}{School of Astronomy and Space Science and Key Laboratory of Modern Astronomy and Astrophysics in Ministry of Education, Nanjing University, Nanjing 210093, China; zhoujl@nju.edu.cn}
\altaffiltext{2}{Key Laboratory of Optical Astronomy, National Astronomical Observatories, Chinese Academy of Sciences, Beijing 100012, China; zhouxu@bao.ac.cn}
\altaffiltext{3}{Department of Astronomy, Beijing Normal University, Beijing 100875, China}
\altaffiltext{4}{Purple Mountain Observatory, Chinese Academy of Sciences, Nanjing 210008, China}
\altaffiltext{5}{School of Physics, University of New South Wales, NSW 2052, Australia}
\altaffiltext{6}{Nanjing Institute of Astronomical Optics and Technology, Nanjing 210042, China }
\altaffiltext{7}{Australian Astronomical Observatory, NSW 1710, Australia }
\altaffiltext{8}{Polar Research Institute of China, Pudong, Shanghai 200136, China}
\altaffiltext{9}{Department of Astronomy and Astrophysics, University of California, Santa Cruz, CA 95064, USA}

%% Mark off your abstract in the ``abstract'' environment. In the manuscript
%% style, abstract will output a Received/Accepted line after the
%% title and affiliation information. No date will appear since the author
%% does not have this information. The dates will be filled in by the
%% editorial office after submission.

\begin{abstract}

The Chinese Small Telescope ARray (CSTAR) is the first telescope facility built at Dome A, Antarctica.
During the 2008 observing season, the installation provided long-baseline and high-cadence photometric observations in the $i$-band for 18,145 targets within $20\,{\rm deg}^2$ CSTAR field around the South Celestial Pole for the purpose of monitoring the astronomical observing quality of Dome A and detecting various types of photometric variability. Using sensitive and robust detection methods, we discover 274 potential variables from this data set, 83 of which are new discoveries. We characterize most of them, providing the periods, amplitudes and classes of variability. The catalog of all these variables is presented along with the discussion of their statistical properties.

\end{abstract}

%% Keywords should appear after the \end{abstract} command. The uncommented
%% example has been keyed in ApJ style. See the instructions to authors
%% for the journal to which you are submitting your paper to determine
%% what keyword punctuation is appropriate.

%\keywords{clusters: globular, peanut---bosons: bozos}
\keywords{binaries: eclipsing --- catalogs --- methods: data analysis --- stars: variables: general --- surveys --- techniques: photometric}

%% From the front matter, we move on to the body of the paper.
%% In the first two sections, notice the use of the natbib \citep
%% and \citet commands to identify citations.  The citations are
%% tied to the reference list via symbolic KEYs. The KEY corresponds
%% to the KEY in the \bibitem in the reference list below. We have
%% chosen the first three characters of the first author's name plus
%% the last two numeral of the year of publication as our KEY for
%% each reference.

\section{INTRODUCTION}

The study of variable stars has long been an essential part of astronomical research and is the mainstay for understanding stellar properties as well as stellar formation and evolution. Variable stars play a crucial role in such astrophysical pursuits as the age of Universe, the cosmological distance scale, the composition of the interstellar medium and the behavior of the expanding Universe.

Large telescopes are not suitable for studies of low-amplitude variations at high cadence, due to limitations of observing time \citep{swift2014}, lack of proper instrumentation, and/or insufficient photometric precision \citep{Tonry2005}. In recent years, there has been a rapid progress in the longitude-distributed (e.g., HATNet: Bakos et al. 2004; HATSouth: Bakos et al.2013) and space-based (e.g., \textit{Kepler}: Borucki et al. 2010; \textit{CoRoT}: Baglin et al.2006) transiting surveys providing an enormous amount of high-precision time-resolved photometric data, resulting in a rapidly increasing number of variability detections that are collected by variable star catalogs such as the Variable Star Index\setcounter{footnote}{9}\footnote{http://www.aavso.org/vsx/} (VSX) or the General Catalog of Variable Stars (GCVS, Samus et al. 2012). In addition to these surveys, the variable sky can be efficiently explored by the Antarctic photometric survey.

The extremely cold, dry, steady, transparent, and dark Antarctic winter skies provide favorable conditions for a diverse and extensive range of astronomical observations, including photometric variability detection \citep[and references therein]{Burton2010}; The long polar night in Antarctica greatly facilitates the detection of low-amplitude variables with relatively long period which require continuous photometric monitoring. Furthermore, the low levels of atmospheric turbulence at Antarctica results in a decrease in scintillation noise, leading to superior photometric precision \citep{Kenyon2006}. The pre-eminent conditions for photometric observations at Antarctic Plateau have been utilized and quantified by the previous observing facilities conducted at different Antarctic sites: SPOT \citep{Taylor1988} at the Amundsen-Scott South Pole Station; the small-IRAIT \citep{Tosti2006}, ASTEP-South \citep{Crouzet2010}, and ASTEP-400 \citep{Daban2010} at Dome C Concordia Station.

Dome A, located $1,200\,{\rm km}$ inland on the Antarctic Plateau, is thought to be the coldest place on Earth. At $4,093\,{\rm m}$ Dome A is also the highest ice feature of Antarctica. An analysis carried out by \citet{Saunders2009} who considered the weather, precipitable water vapor, the boundary layer, the free atmosphere, aurorae, airglow, thermal sky emission, and surface temperature led them to conclude that Dome A may be the best site on Earth.

\begin{figure}
\epsscale{0.85}
\plotone{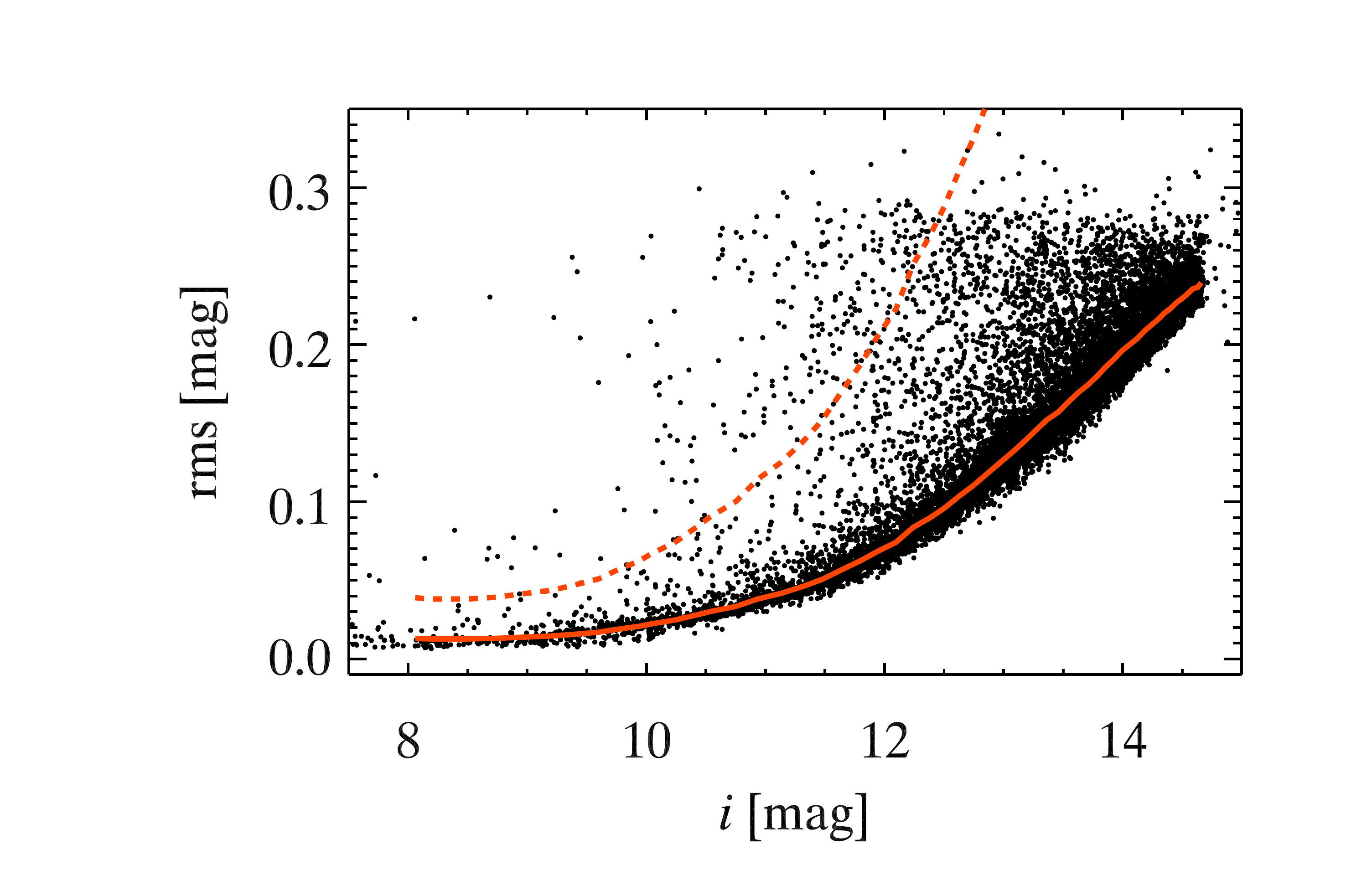}
\caption{Photometric quality of the CSTAR 2008 data set. The standard deviation (20 second or 30 second sampling; 158 day time scale) of each CSTAR light curve is plotted as a fuction of their median $i$ magnitudes. The solid orange line represents the trend of this distribution. Objects above $3\,\sigma$ threshold (the dashed orange line) are tagged as variable candidates.
\label{fig1}}
\end{figure}

With the aim of taking advantage of and quantifying these favorable astronomical observing conditions at Dome A, the Chinese Small Telescope ARray (CSTAR) was shipped and established there in 2008 January. In the same year, approximately 300,000 \textit{i}-band photometric images were collected during nearly continuous observations for more than four months, resulting in a high-precision catalog (see Figure~\ref{fig1}) of 18,145 stars in a field centered on the South Celestial Pole \citep{Zhou2010a, Wang2012, Wang2014a}, which makes these data an excellent source for detection of various types of photometric variability.

A first characterization of the stellar variability in the CSTAR field has been published by \citet{Wanglz2011, Wanglz2013}. In this paper we present the more fruitful results of a search for variable stars in this field based on the higher-precision light curves, obtained from independent analysis of the CSTAR data set from the observations in 2008.

The layout of the paper is as follows. The CSTAR telescope system setup, the observational strategy and data reduction processes are briefly described in Section 2. In Section 3, we detail the techniques of variable-star detection and period determination. In Section 4, we present and discuss the results of variable-star search in the CSTAR field. Lastly, the work is summarized in Section 5.

\section{INSTRUMENT, OBSERVATIONS AND PREVIOUS DATA PROCESSING}

\subsection{Instrument}
CSTAR, controlled from the PLATO autonomous observatory \citep{Lawrence2009, Yang2009}, consists of four fixed, co-aligned $14.5\,{\rm cm}$  (effective aperture of $10\,\rm{cm}$) telescopes, each with a different optical filter in SDSS bands: \textit{r, g, i} and \textit{open}.
Each telescope is equipped with a $1\rm K \times 1 \rm K$ Andor DV 435 frame transfer CCD camera with a pixel size of $13\,{\rm \mu m}$ and an angular resolution of $15\,\rm{arcsec}\,\rm{pixel}^{-1}$, yielding a $4^\circ.5 \times 4^\circ.5$ wide field of view (FOV) around the South Celestial Pole. Details of the CSTAR facility can be found in \citet{Yuan2008} and \citet{Zhou2010b}.

\subsection{Observations}

CSTAR, the first photometric telescope to enter operation at Dome A, was successfully deployed there in 2008 January, and operated for the subsequent four winters until it was retrieved in 2012 to be repurposed. The data set analyzed in this work was collected from 2008 March 4 to August 8. In this observing season, about $1728\,\rm{hr}$ observations provided some $0.3\,{\rm million}$ $i$-band frames for 18,145 stars with exposure times of $20\,\rm s$ or $30\,\rm s$. A detailed description of the CSTAR observations in 2008 is given in \citet{Zhou2010a}.

\subsection{Previous Data Reduction}
To achieve mmag photometric precision for the bright CSTAR objects, the data set was calibrated and reduced using a custom reduction pipeline as described in detail in \citet{Zhou2010a} and \citet{Wang2012, Wang2014a}. Here only the main factors to be considered when reducing the CSTAR data are briefly reviewed.

After the bias and the flat field were corrected, aperture photometry was performed on all CSTAR frames. 
Using the 48 bright local calibrators, the CSTAR instrumental magnitudes were then calibrated to the SDSS $i$. 
Instead of registering every CSTAR photometric catalog to an external astrometric reference catalog, we generated light curves of 18,145 point sources in the fixed CSTAR field by matching all CSTAR photometric catalogs to a master catalog using a triangle matching algorithm \citep{valdes1995}. The master catalog, constructed from 10,000 CSTAR frames taken in the best photometric conditions, was astrometrically registered to the USNO-B catalog \citep{Monet2003} by only solving for the frame center and rotation angle, producing relatively large astrometric residuals (The coordinates of identified variables in this study are provided by cross matching against 2MASS catalog : Skrutskie et al. 2006.  The original CSTAR coordinates are retained only as the index of the released light curves: Zhou et al. 2010a, Wang et al. 2012, 2014a).
After that, the first version of the CSTAR photometric catalog, detailed in \citet{Zhou2010a}, was constructed.

For detection of low-amplitude photometric variability, including planetary transits, the first version of the CSTAR photometric products were further refined by employing corrections for additional systematic errors, as outlined below.

For mmag photometry, uneven atmospheric extinction across the large CSTAR FOV ($4.5^{\circ} \times 4.5^{\circ}$), especially under bad observing conditions, cannot be ignored; This was modeled and corrected by comparing each CSTAR photometric frame to a master frame. For details on how this refinement was achieved, see \citet{Wang2012} .

The residual of the flat-field correction shows up as daily systematic variations in stellar flux during the diurnal motion of the stars on the static CSTAR optical system. It was effectively corrected by comparing each target object to a bright and constant reference star in the nearby diurnal path. We refer the reader to \citet{Wang2014a} for more details.

%Since CSTAR is a static telescope and fixed to point at the South Celestial Pole, star images move clockwise on the CCD due to diurnal motion. Ghost images, located in symmetrical position of the CCD, move counterclockwise. Because of this, ghost images move and contaminate the photometry of stars. The significant contamination arising from the ghost images, detailed in \citet{Meng2013}, was studied and corrected.

\section{VARIABLE STAR DETECTION}
In this section, we concentrate on the procedures for the variable star detection, beginning with a review of the final photometric precision of the CSTAR data, continuing with a description of the methods used to sift the variables, and finishing with a discussion of the robust techniques adopted to eliminate the false-positive detections.

\subsection{Photometric Precision}
The resulting light curves from the ensemble photometry stage of our pipeline typically reach a precision of $\sim 0.004\,\rm{mag}$ at $20\,\rm s$ or $30\,\rm s$ cadence for the brightest non-saturated stars ($i=7.5$), rising to $\sim 0.02\,\rm{mag}$ at $i=12$. The distribution of the standard deviation of light curves is shown in Figure~\ref{fig1}. Each point represents the rms of a $20\,\rm s$ or $30\,\rm s$ sampled light curve with more than one hundred-day observations.

\subsection{Detection of Variables}

\subsubsection{Detection of Variables in General}
In this section, we identify possible photometric variability from the CSTAR data set in two steps. The first involves the relation between
the light-curve standard deviations and their median $i$ magnitudes as shown in Figure~\ref{fig1}.
Naturally, the variable stars are expected to present higher deviations than non-variable stars with the
same brightness. For that reason an object is tagged as a variable candidate if its standard deviation is three times higher than the typical value at its magnitude.

\begin{figure}
\epsscale{0.85}
\plotone{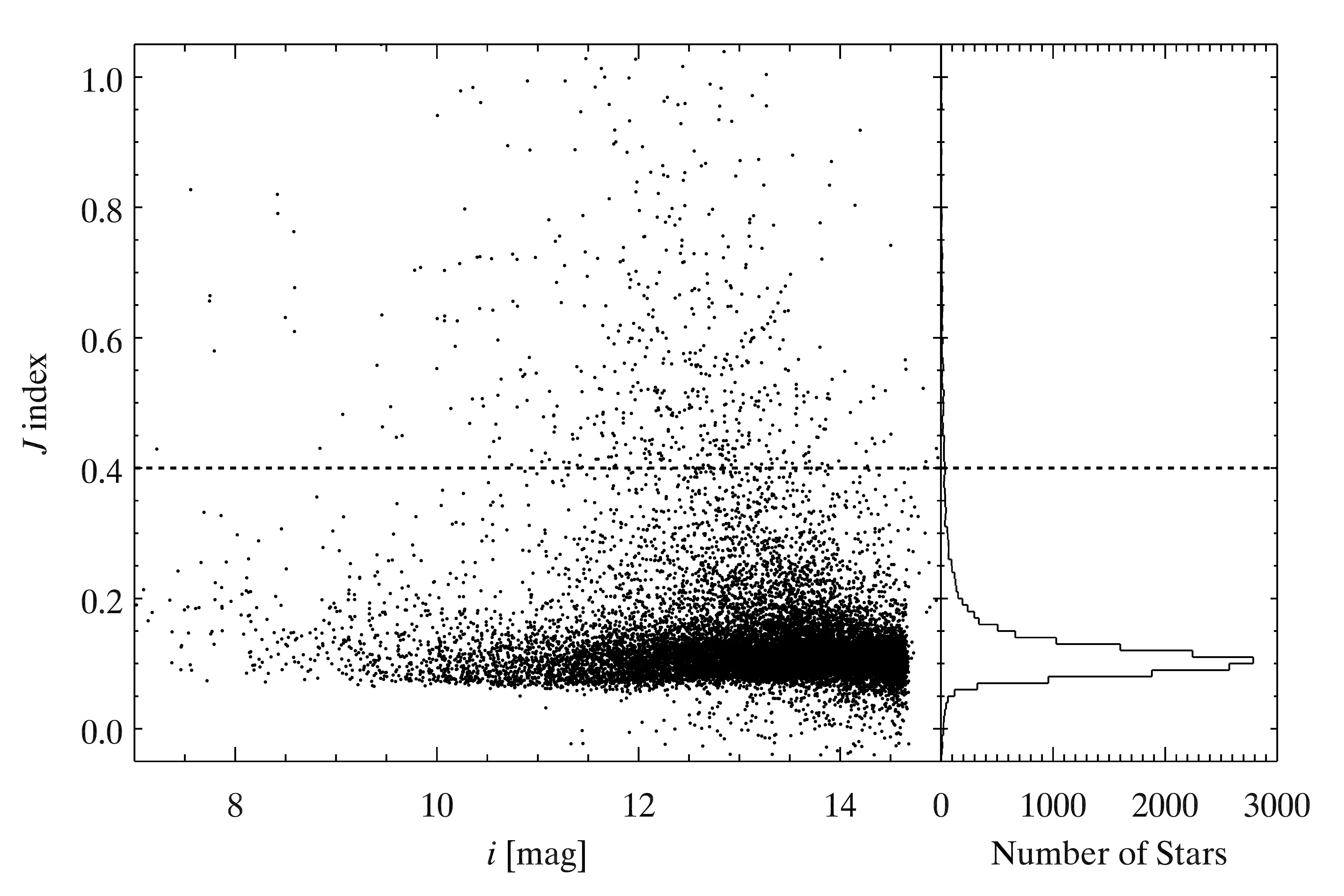}
\caption{Distribution of the Stetson variability $J$ index for all the objects in the CSTAR field. The dashed line marks our threshold for variability detection, $J=0.4$.
\label{fig2}}
\end{figure}

As pointed out by \citet{Rose2007}, this method is not sensitive to low-amplitude variables. Moreover, remaining systematic effects or even several problematic data points in the light curve will result in a higher deviation, which can give rise to false-positive variability detection. To further identify variable stars, we employ the Stetson variability index $J$ defined by \citet{Stetson1996} and modified by \citet{Zhang2003} to our data set. The $J$ statistic is given as
\begin{equation}
J = \frac{\sum_{k=1}^{n} \omega_k {\rm sgn}(P_k) \sqrt{|P_k|}}{\sum_{k=1}^{n} \omega_k},
\end{equation}
where $\omega_k$ is the time-related weighting factor assigned to the $k^{\rm th}$ pair of measurements, calculated by
\begin{equation}
\omega_k = {\rm exp}(-\frac{\Delta t_k}{\overline{\Delta t}}),
\end{equation}
where $\Delta t_k$ is the time interval for the $k^{\rm th}$ pair of measurements and $\overline{\Delta t}$ is the median time interval for all pairs of measurements.
The expression $P_k$ is the product of the normalized residual for the $k^{\rm th}$ pair of measurements and is given by
\begin{equation}
P_k = \frac{n}{n-1}(\frac{m_{1,k}-\overline{m}}{\sigma_{1,k}})(\frac{m_{2,k}-\overline{m}}{\sigma_{2,k}}),
\end{equation}
where $m_{1,k}$ and $m_{2,k}$ are the first and second magnitudes of the $k^{\rm th}$ pair of measurements, $\sigma_{1,k}$ and $\sigma_{2,k}$ denote the propagated errors associated with these measurements. $\overline m$ is the median magnitude, and $n$ is the number of observational pairs.
The Stetson $J$ index is small when the normalized magnitude residuals $P_k$ are uncorrelated, as in the case of non-variable stars, even those with high values of standard deviation.  Real variable stars have correlated magnitude measurements across pairs of subsequent observations, which will increase the Stetson $J$ index for such sources. It is therefore particularly reliable when checking for correlated variations in subsequent measurements. Figure~\ref{fig2} presents the distribution of the variability index $J$ for all target objects in the CSTAR data set. The empirical limiting value of $J=0.4$ is applied to distinguish possible variable candidates from non-variable stars.

\subsubsection{Further Detection of Periodic Variables}
\begin{figure}
\epsscale{0.85}
\plotone{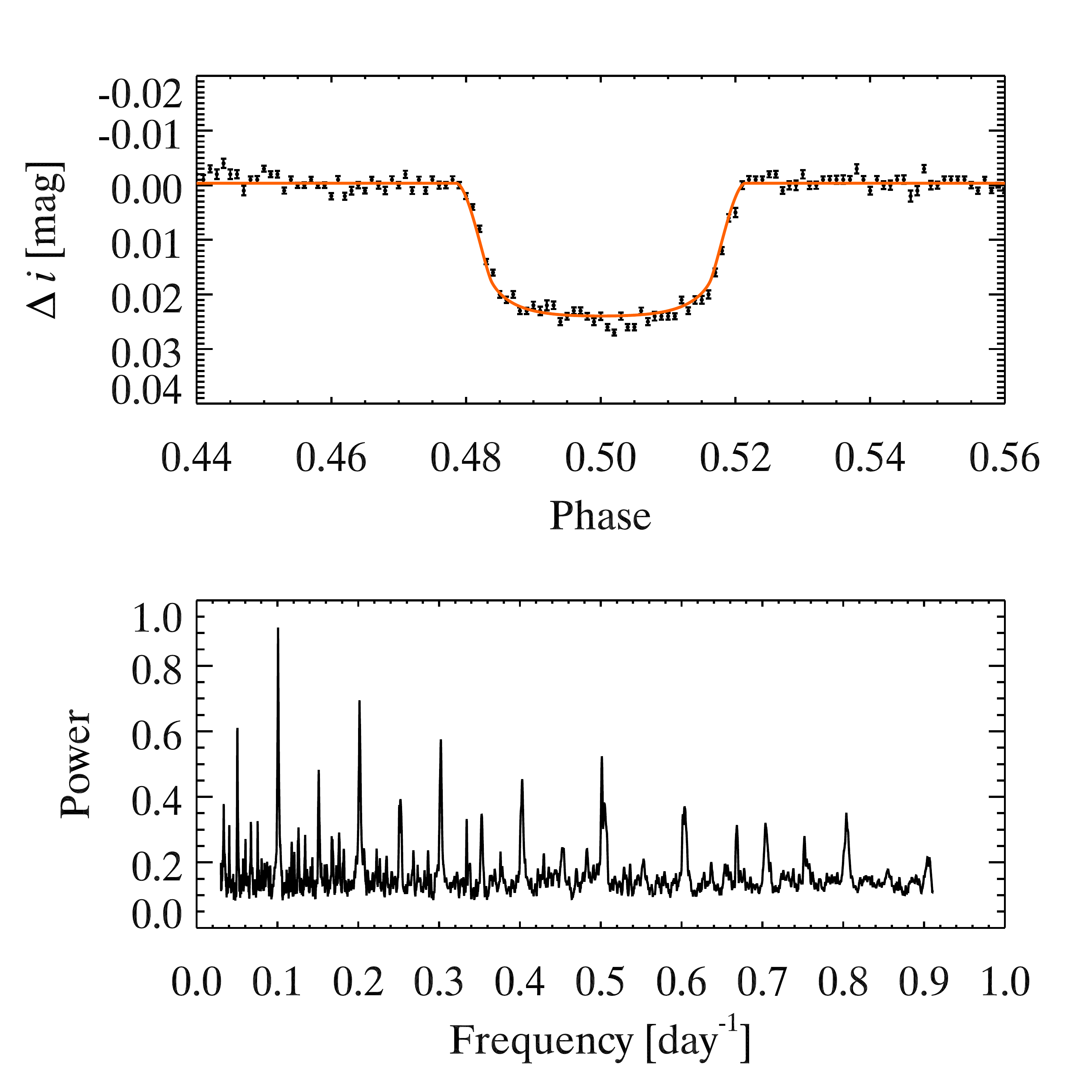}
\caption{Example of a clear low-amplitude variable (CSTAR J183051.60-884322.6) in the CSTAR data set. It cannot be readily identified through the statistical analysis of mag-rms relation or Stetson variability $J$ index due to both low ${\rm rms}=0.029$ and low $J\,\,{\rm index}=0.088$  of this star. For that reason, use of AoV and BLS periodic analysis is motivated by the possibility that low-amplitude variables (as in this case) may have potential periodicities which could be reflected in their periodograms (lower panel).
\label{fig3}}
\end{figure}

An example of a clear transit event, obtained by the transit search \citep{Wang2014b} in the CSTAR data set, is shown in Figure~\ref{fig3}. The low $J=0.088$ and $\rm{rms}=0.029$ values of this star indicate that both methods described in section 3.2.1 are not effective approaches for detecting low-amplitude periodic variables. In order to maximize the detection yield and to search for periodicities, we perform a further analysis of our data set especially for periodic variables.

\begin{figure}
\epsscale{0.85}
\plotone{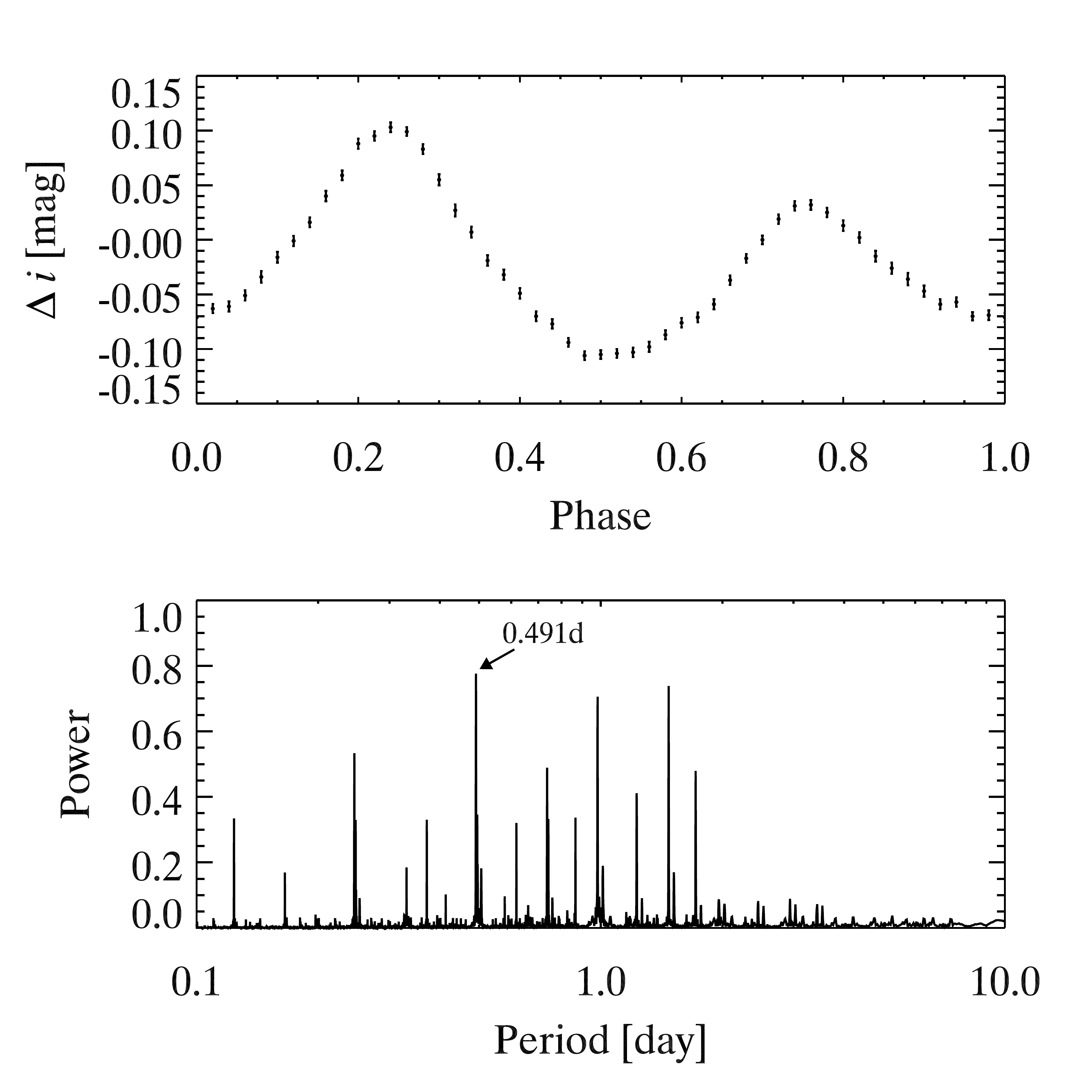}
\caption{Example for one of the periodic variable stars (CSTAR J061948.66-872043.4) detected in this study. The top panel shows the $i$-band light curve phased on its period of 0.491 days, with AoV periodogram plotted below. The period of this object is indicated by the arrow.
\label{fig4}}
\end{figure}

Two methods are applied to search for periodic signals among all the CSTAR objects. The analysis of variance (AoV) method, introduced by \citet{Schwarzenberg1996}, is applied as the first step. In this method, after the light curve is phased and binned with a series of trial periods, the best period is determined to minimize the ratio of the intra-bin to the overall inter-bin variances. Although it is a time-consuming process, the AoV statistic is applied with $N=7$ harmonics to all the CSTAR objects to obtain power spectra between 0.01 and $100\,\rm{days}$. Light curves which show significant AoV statistics (detection threshold $>$ 6) are folded with their respective frequencies and then inspected visually before accepted as periodic variables. Figure~\ref{fig4} shows as an example the phased light curve of a randomly selected periodic variable found in this manner.

Since this method is not optimal for detecting transit events, a more sophisticated transit-sifting algorithm (BLS, Kov{\'a}cs et al. 2002) is applied to the CSTAR data set to identify three transit-like events (see Figure~\ref{fig3} for example).

\subsection{Data Validation}
Since the detection products are inevitably affected by the remaining systematic effects presented in the CSTAR data set, additional tests are performed to distinguish spurious signals from the true stellar variability. The variable candidates which meet any of the following criteria are discarded.

\begin{itemize}
  \item \textit{Frequencies with Poor Phase Coverage.} Gaps or clumpy data points in the phased light curve would lead to aliasing. A visual inspection is made of each detected variable candidate. Surviving variable candidates are required to have smoothly sampled phased light curves.

  \item \textit{Periods at Known Aliases.} Both AoV and BLS period-search algorithms suffer from aliasing originating from the remaining systematic errors in the CSTAR data set. It generates false period peaks at frequencies associated with $1\,{\rm d}$ and at some other commonly occurring frequencies. To minimize the number of false-positive detections, objects with detected periods close to known aliases are excluded.

  \item \textit{Photometric contamination}. False variablity may result from non-variable objects contaminated by the nearby variables. This kind of spurious variables can be eliminated by comparing their light curves to the light curves of nearby sources that can be resolved by the CSTAR images. Note the large pixel scale ($15\,\rm{arcsec}\,\rm{pixel}^{-1}$) of CSTAR would inadvertently leave unresolved blending in the data set, which reqiure follow-up time-series photometry with instrumentation giving high spatial resolution to confirm their variability.

      In addition, in case of uncertainty, the related images are inspected by eye. A hot pixel or cosmic ray can affect the star or sky value used for photometry. The bright wings of saturated stars or a satellite track can also seriously impact the measurements. After detailed visual inspection, 274 objects which show robust variability are finally selected as the reliable variables in the CSTAR field.

\end{itemize}

\section{RESULT AND DISCUSSION}

\subsection{Result and Statistical Discussion}

About 18,145 objects down to $i=14.8$ are used to detect variability in 20 square degrees of the CSTAR FOV. The variability-searching process finally yields 274 ($\sim 1.51$ percent of the total objects) variables, including 221 with clear periodicity. All these objects along with their detailed information are presented in Table~\ref{table1} and Table~\ref{table2}. The stellar identifer is of the form `${\rm CSTAR\,\,J}hhmmss.ss+ddmmss.s$', based on their coordinates from the first release of the CSTAR photometric data set \citep{Zhou2010a}.

As a summary of our findings for the CSTAR project, histograms of the amplitude, $J$ index, magnitude as well as determined period for the detected variable stars are shown in Figures 5 to 8.

\begin{figure}
\epsscale{0.85}
\plotone{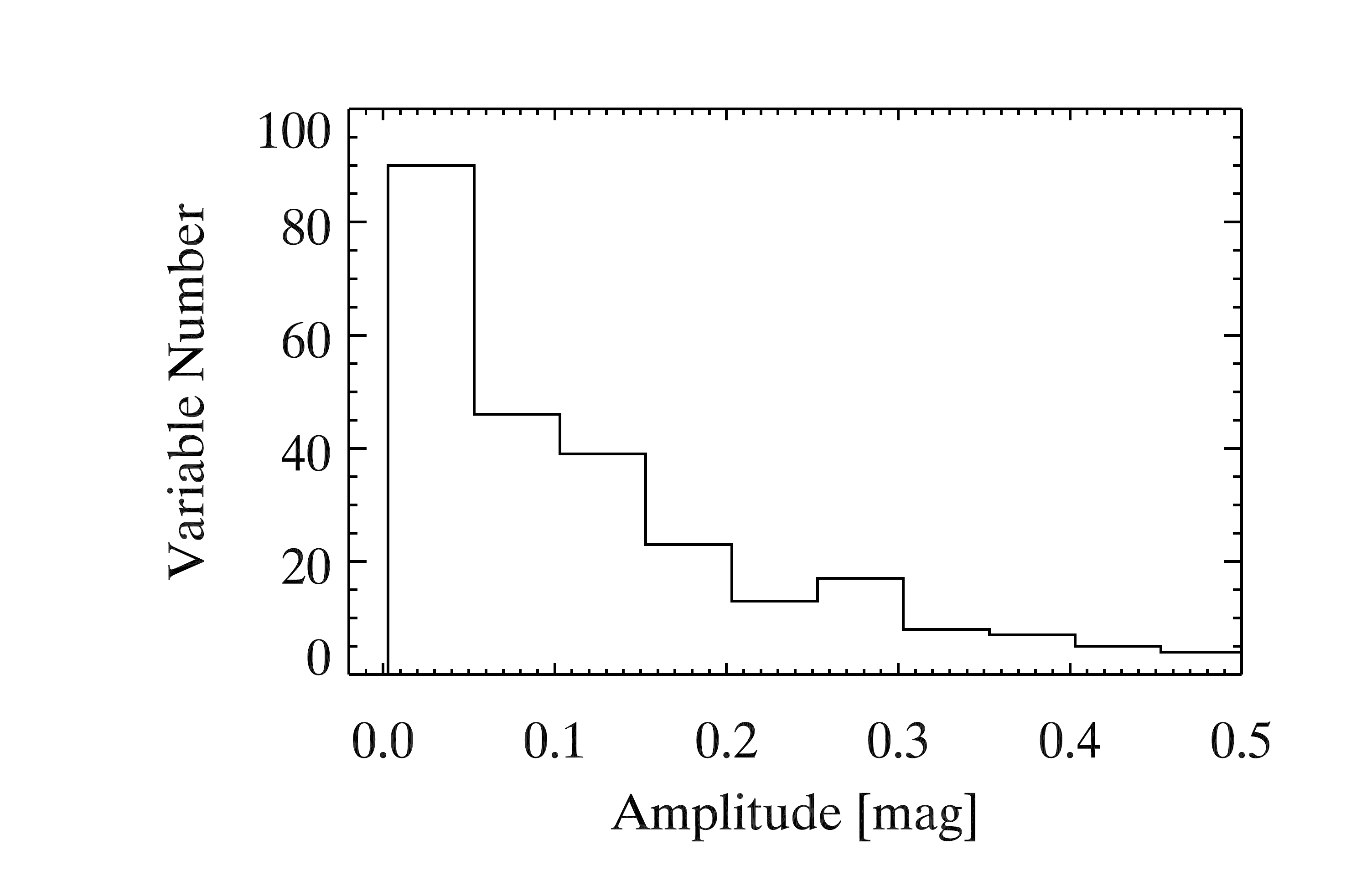}
\caption{Histogram of amplitude for all the variables found in the CSTAR 2008 data set. Note a rough decline from small to large amplitude.
\label{fig5}}
\end{figure}

The amplitude distribution (Figure~\ref{fig5}) for the identified variables in the CSTAR data set yields a rapid falloff at larger amplitude. A significant number of small amplitude (${\rm amplitude}\leq 0.05$) variables is detected. Not surprisingly, the fraction of stars that are significantly variable is less than the fraction of stars with small amplitude variability.

\begin{figure}
\epsscale{0.85}
\plotone{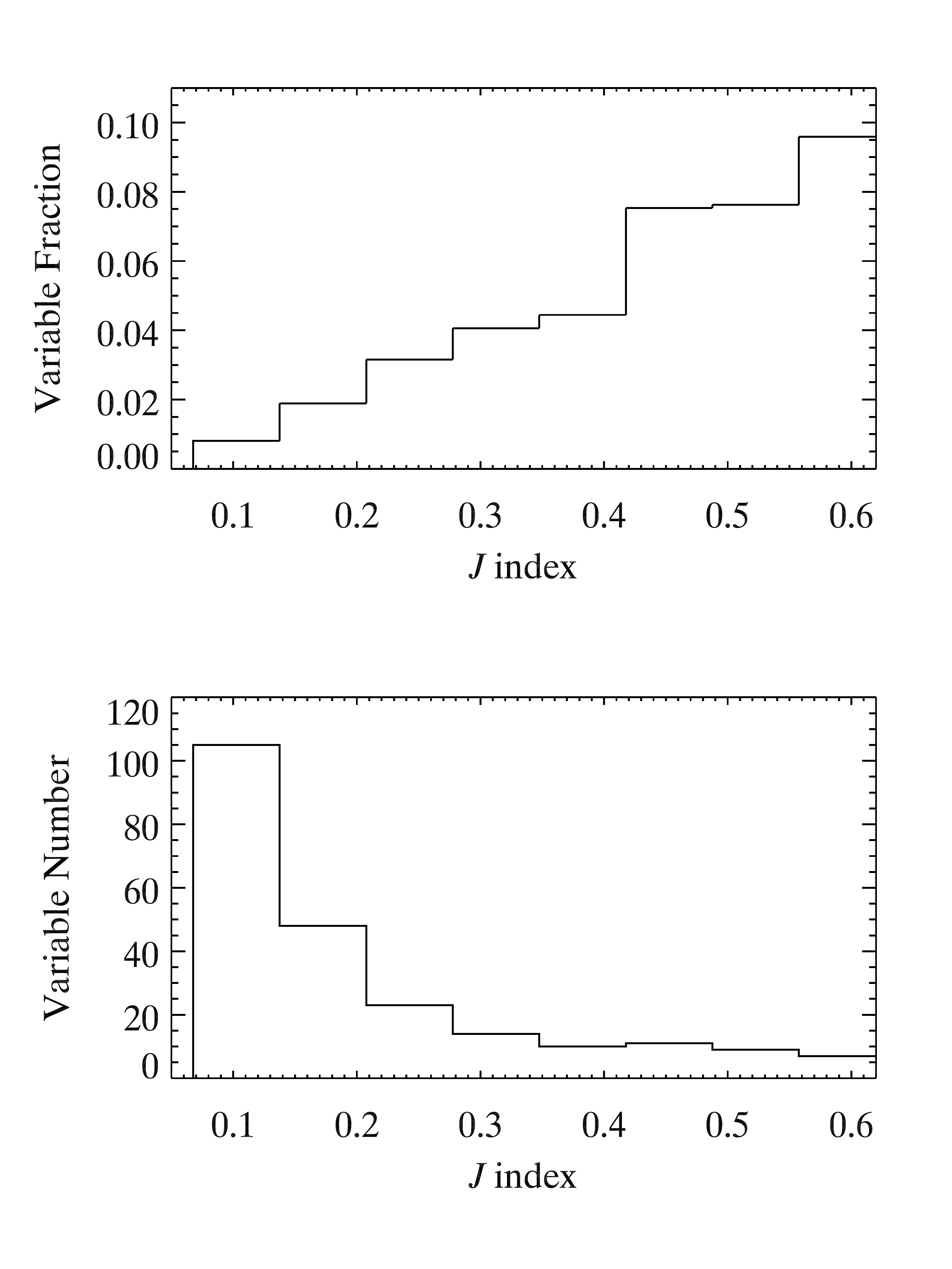}
\caption{(top) Fraction of variable stars as a function of Stetson variability $J$ index.
Although the variability-detection fraction increases with the $J$ index,
a number of identified variables with relatively lower $J$ index indicates that the Stetson variability $J$ index statistic alone is not an effective variable-selection criterion. The multiple detection techniques are required to ensure the completeness of the detection.
(bottom) Distribution in the number of variable stars as a function of $J$ index. Similar to Figure~\ref{fig5}, we note an expected decline from small to large $J$.
\label{fig6}}
\end{figure}

Although the variability-detection fraction increases with the $J$ index (upper panel in Figure~\ref{fig6}), a number of identified variables with relatively lower $J$ index (bottom panel in Figure~\ref{fig6}) indicates that the Stetson variability $J$ index statistic alone is not an effective variable-selection criterion. That is the reason that the multiple detection techniques are adopted in this paper, which significantly increase the efficiency of detection.

\begin{figure}
\epsscale{0.85}
\plotone{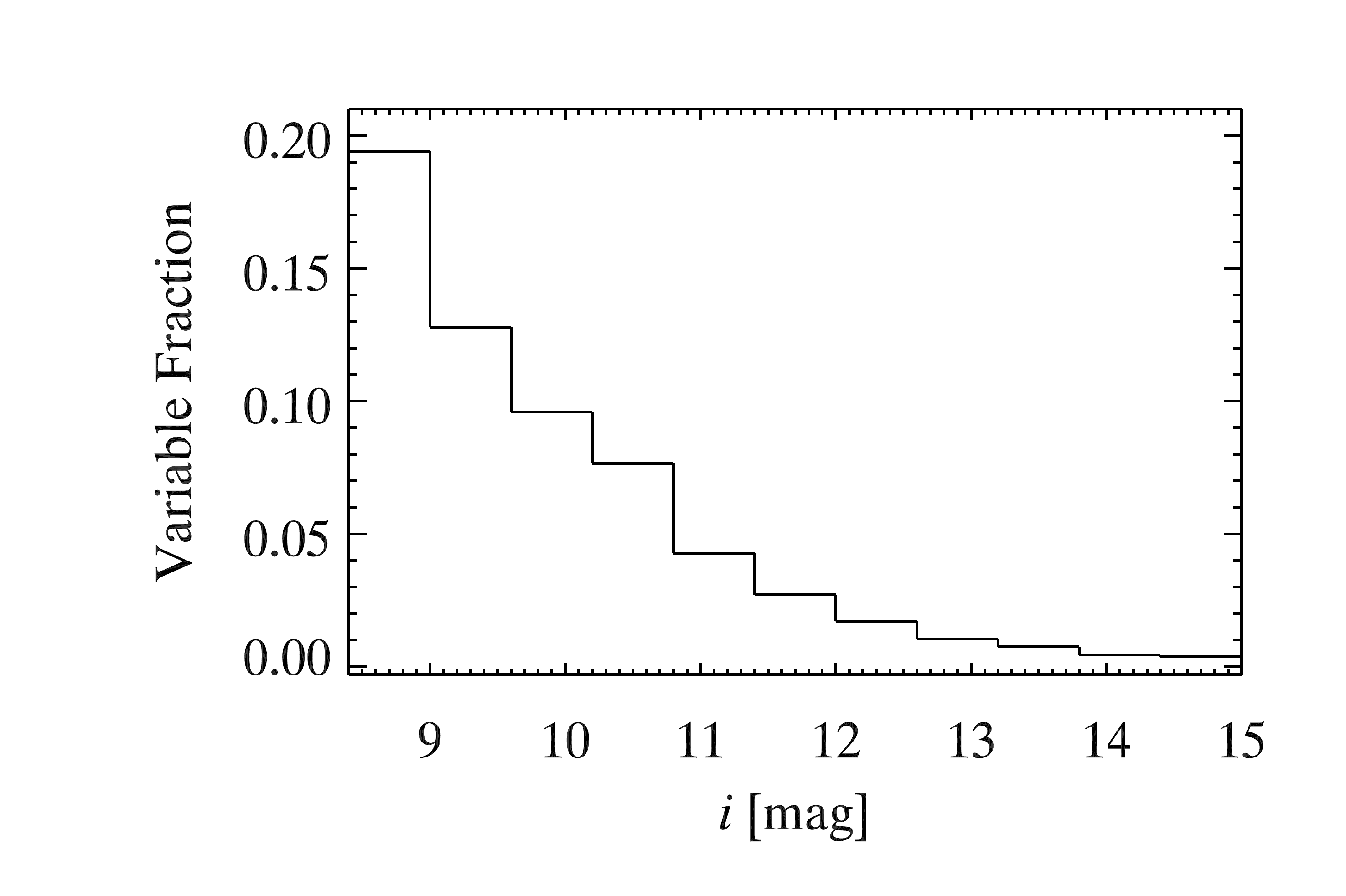}
\caption{Fractional distribution of magnitude for the identified variables in the CSTAR field.
\label{fig7}}
\end{figure}

The increasing photometric noise in the faint end would lead to a fall in variability-detection fraction. This selection bias is reflected clearly in the fractional distribution of the identified variables as a function of their $i$ magnitude (Figure~\ref{fig7}).

\begin{figure}
\epsscale{0.85}
\plotone{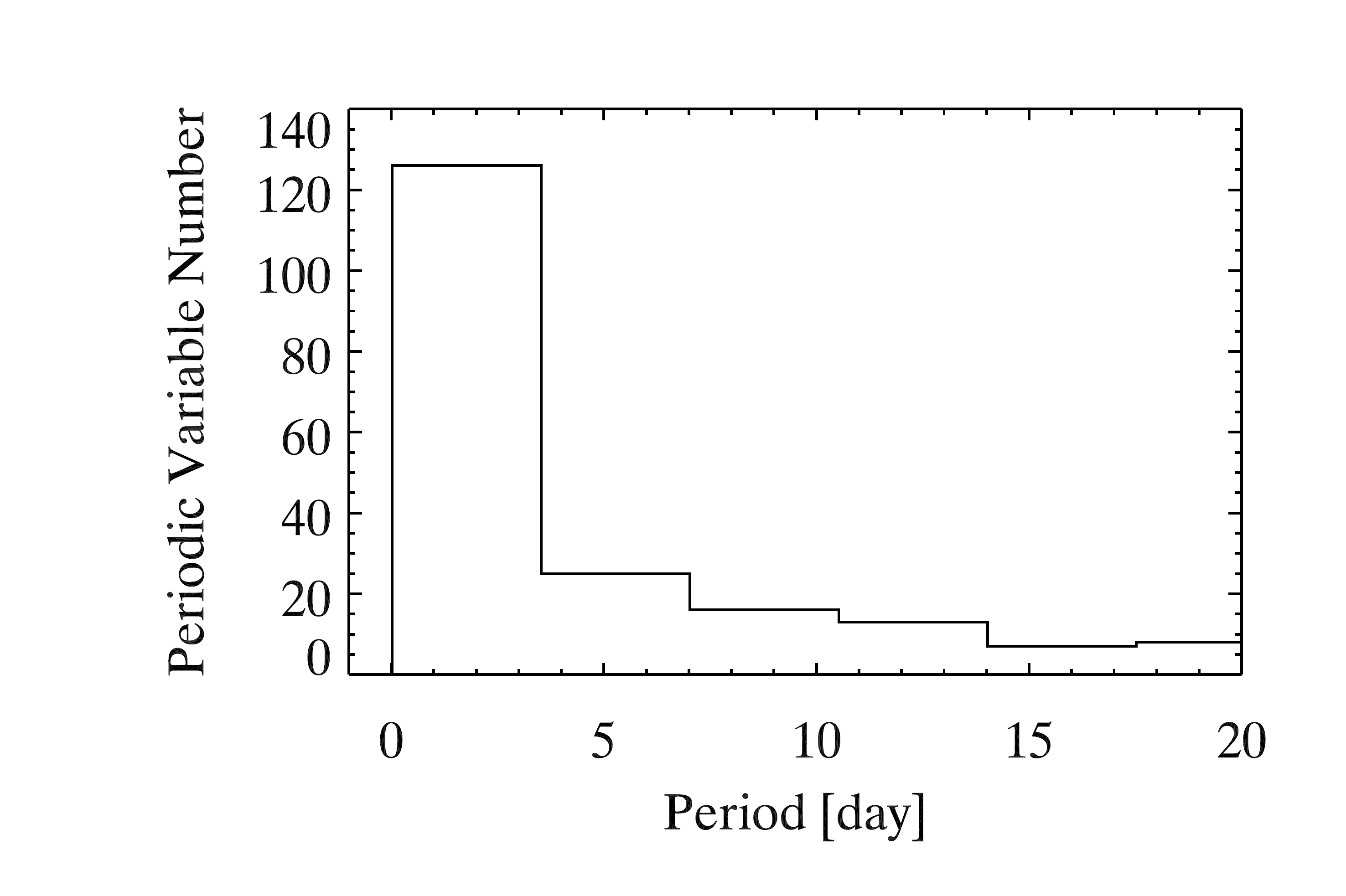}
\caption{Period distribution of the detected periodic variables in the CSTAR 2008 data set. We note a rapid falloff at longer period and a peak in the period of $< 3.5\,\rm{d}$. Although the efficiency of detecting short-term variables is significantly higher, thanks to more than four-month high-duty-cycle observation during the Antarctic winter, the longest period of detected variables in the CSTAR 2008 data set is up to $ 88.39\,\rm{d}$.
\label{fig8}}
\end{figure}

We see that the period histogram (Figure~\ref{fig8}) for the detected periodic variables yields a somewhat expected distribution. The distribution peaks at slightly less than $3.5\,\rm{d}$, and the shorter periods are 5 times more prevalent.  Our data set is not sensitive to longer periods, partly due to (1) limitation of the observational time baseline, and (2) the intrinsic frequency of these short-term variables in the entire population of the variables.

\subsection{Comparison with Previous Work}

%To confirm our process of variability and their periodicities detections as well as the quality of the photometric conditions in Dome A, detected variables from the %CSTAR data set have been cross-matched with the GCVS catalog via SIMBAD and AAVSO catalog. We have recovered ?? of ?? previously variables in these catalogs.
%?? of missing variables, with $i<$, are saturated in the CSTAR data.

CSTAR data collected during both 2008 and 2010 observing season have been independently analyzed and used to detect the variables by \citet{Wanglz2011, Wanglz2013}; It provides a great opportunity to compare and contrast the scientific results of the two groups in detail.

The identified variables in this study are cross-matched with the previous variable catalog \citep{Wanglz2011, Wanglz2013} from the same data set. We recover 191 of 224 previously known variables. In Table~\ref{table1}, we summarize our recovery attempts in detail.
The periods determined by us are generally in agreement with the periods given by \citet{Wanglz2011, Wanglz2013}. The improved periods for seven stars which show a clear disagreement in the period determination are checked carefully by reviewing their periodograms and light curves.

Moreover, 83 of our variables have no best counterpart in theirs. These objects are considered as new variables and are summarized in Table~\ref{table2}. These objects, including 60 periodic variables (see Figure~\ref{fig9}, phased light curves) and 23 non-periodic or quasi-periodic variables (see Figure~\ref{fig10}, light curves), with quite clear variability match our detection criteria very well, but were not reported by \citet{Wanglz2011, Wanglz2013}.

On the other hand, we are not able to confirm the photometric variability for 33 targets reported as variables by \citet{Wanglz2011, Wanglz2013}. All of these cases are listed in Table~\ref{table3}, and  are carefully inspected. It is likely that four of their variability are originating from systematics with the aliased frequencies. The 29 objects show no significant photometric variability at the limits of our detection sensitivity.

\subsection{Variability Classification}

The variability of stars is usually caused by intrinsic (pulsating, eruptive, cataclysmic) or extrinsic (eclipsing, rotating) factors, or any combination of them.

To assist in the classification of the identified variables in the CSTAR data set, the luminosity classes and spectral types given in the VizieR \citep{ochsenbein2000} are utilized, when available.   
Additionally, the remaining variables are cross-correlated with existing catalogs using a $30''.0$ match radius to provide their $JHK$ magnitudes (2MASS: Skrutskie et al. 2006) and proper motions (PPMXL: Roeser et al. 2010). This allow us to estimate their luminosity classes and spectral types. 
 
 We use the $V-K$ color together with the reduced proper motion (RPM;  Luyten 1922) to separate the main-sequence dwarfs from giants \citep{Street2007, clarkson2007, Hartman2011}.
 Here the ${\rm RPM}_V$, is calculated as 
\begin{equation}
{\rm RPM}_V = V+5{\rm log}_{10}(\mu/1000),
\end{equation}
where the $V$ magnitudes are transformed from their cross-referenced 2MASS $JHK$ magnitudes \citep{Hartman2011}, $\mu$ is the proper motion in units of ${\rm mas\,yr^{-1} }$ taken from the PPMXL catalog \citep{roeser2010}.
\{${\rm RPM}_V$, ($V-K$)\} space, as displayed in Figure~\ref{fig11}, has been shown to be a powerful diagnostic tool in distinguishing dwarfs from giants \citep{Street2007, clarkson2007, Hartman2011}, as the latter lean towards lower values of ${\rm RPM}_V$, and higher $V-K$.
For the variables without recorded spectral classification, The color indices, derived from 2MASS $JHK$ magnitudes, are used to estimate their expected spectral types \citep{Bessell1988}. 
Given the uncertainties inherent in the above analyses which are made from proper motion and color indices alone, these rough estimates are thus not provided in the final variable catalog.
 
Based on these detailed stellar information  together with the noteworthy features (shape, period, and amplitude) of the variable light curves,
all the detected variables in our data set are grouped into the following classes according to the GCVS-based schema \citep{Samus2012}.

Pulsating stars are described by $\delta$ Scuti type (DSCT), RR Lyrae type (RL), $\gamma$ Doradus type (GD), Cepheids (CEP), and $\delta$ Cephei-type (DCEP) variables. Eclipsing binaries are subdivided into Algol type (EA), $\beta$ Lyrae type (EB), and W Ursae Majoris type (EW). Some shallow eclipsing signals are identified as transit-like events (TR); Follow-up studies with these stars are needed to confirm whether or not they might be genuine transiting planets. Rotating variables are categorized within the three major groups: BY Draconis-type variables (BY), $\alpha^2$ Canum Venaticorum variables (ACV), and ellipsoidal variables (ELL). 
We also introduce a category spotted stars (SP) for the cases showing variability characteristic features of the stellar spots, but the precise type of which could not be determined.
In addition, some variables exhibiting unknown features are classified as Irregular variables (IR). This class also includes the cases for which the variable periods appear longer than the observational baseline.

The type of variability assigned to each variable star is presented in Table~\ref{table1} and 2. The statistic overview of the classification are provided in Table~\ref{table4}.

Finally, we cross match our detections with the $ROSAT$ catalog \citep{Voges1999, Voges2000} using a $30''.0$ match radius. A total of 12 periodic variables  are successfully correlated  to the $ROSAT$ X-ray sources, including eight of  44 BY Dra stars, three of 23 ellipsoidal variables, and one of 27 Algol type eclipsing binaries, which all are common X-ray emitters \citep{Norton2007, Christiansen2008, Hartman2011}. The results of the correlation are included in Table~\ref{table1} and 2.

Of the 191 recovered variables, our analysis shows that 71 ($37 \%$) agree with the classification in \citet{Wanglz2011, Wanglz2013}, 108 ($57\%$) previous ungrouped variables are classified, and 12 variables ($6 \%$) disagree with the previous category and are reclassified. We are confident that our classification is more reliable, as it is indicated by not only the noteworthy features of the variable light curves, but also their detailed stellar information.

\section{CONCLUSIONS}
In 2008 polar night, more than four months of high-duty-cycle photometric observations with the Antarctic CSTAR telescope provided long-baseline, high-cadence light curves of 18,145 stars in a $20\,\rm{deg}^2$ field centered at the South Celestial Pole.

From this data set we present a catalog of 274 stars exhibiting clear photometric variability, including 83 new variables and 191 already known variables, along with the statistical properties and classification of them. Of all these variables, 58 are eclipsing binaries, 163 are other type periodic variables, and 53 are non-periodic or quasi-periodic variables. It is expected that our knowledge of variables will continue to improve when the CSTAR 2008 data are combined with the multi-band photometric data of subsequent years.

These detections show the favorable quality of Dome A to carry out continuous and long-duration photometric observations, serving a precursor in advance of future photometric surveys conducted at Dome A such as AST3 \citep{cui2008} and KDUST \citep{zhao2011}.

In addition, all of photometric data products, including CSTAR 2008 catalog and light curves for both already known and newly discovered variables, are available online\setcounter{footnote}{10}\footnote{http://explore.china-vo.org/.}.

\acknowledgments
We thank the anonymous referee for the insightful suggestions that greatly improved this manuscript.
This research is supported by
the National Basic Research Program of China (Nos. 2013CB834900, 2014CB845704, 2013CB834902, 2014CB845700, and 2014CB845702);
the National Natural Science Foundation of China under grant Nos. 10925313, 11333002, 11433005, 11073032, 11003010, 11373033, 11373035, 11203034, and 11203031;
the Strategic Priority Research Program: The Emergence of Cosmological Structures of the Chinese Academy of Sciences (Grant No. XDB09000000);
the Main Direction Program of Knowledge Innovation of Chinese Academy of Sciences (No. KJCX2-EW-T06);
the 985 project of Nanjing University and Superiority Discipline Construction Project of Jiangsu Province;
the joint fund of Astronomy of the National Nature Science Foundation of China and the Chinese Academy of Science, under grant Nos. U1231113 and U1231202;
the Natural Science Foundation for the Youth of Jiangsu Province (No. BK20130547);
and the Jiangsu Province Innovation for Ph.D candidate (No. KYZZ$\_$0030).

%%

%% The reference list follows the main body and any appendices.
%% Use LaTeX's thebibliography environment to mark up your reference list.
%% Note \begin{thebibliography} is followed by an empty set of
%% curly braces.  If you forget this, LaTeX will generate the error
%% "Perhaps a missing \item?".
%%
%% thebibliography produces citations in the text using \bibitem-\cite
%% cross-referencing. Each reference is preceded by a
%% \bibitem command that defines in curly braces the KEY that corresponds
%% to the KEY in the \cite commands (see the first section above).
%% Make sure that you provide a unique KEY for every \bibitem or else the
%% paper will not LaTeX. The square brackets should contain
%% the citation text that LaTeX will insert in
%% place of the \cite commands.

%% We have used macros to produce journal name abbreviations.
%% AASTeX provides a number of these for the more frequently-cited journals.
%% See the Author Guide for a list of them.

%% Note that the style of the \bibitem labels (in []) is slightly
%% different from previous examples.  The natbib system solves a host
%% of citation expression problems, but it is necessary to clearly
%% delimit the year from the author name used in the citation.
%% See the natbib documentation for more details and options.

\clearpage

% Use the figure environment and \plotone or \plottwo to include
% figures and captions in your electronic submission.

\clearpage

\begin{figure}
\epsscale{0.75}
\plotone{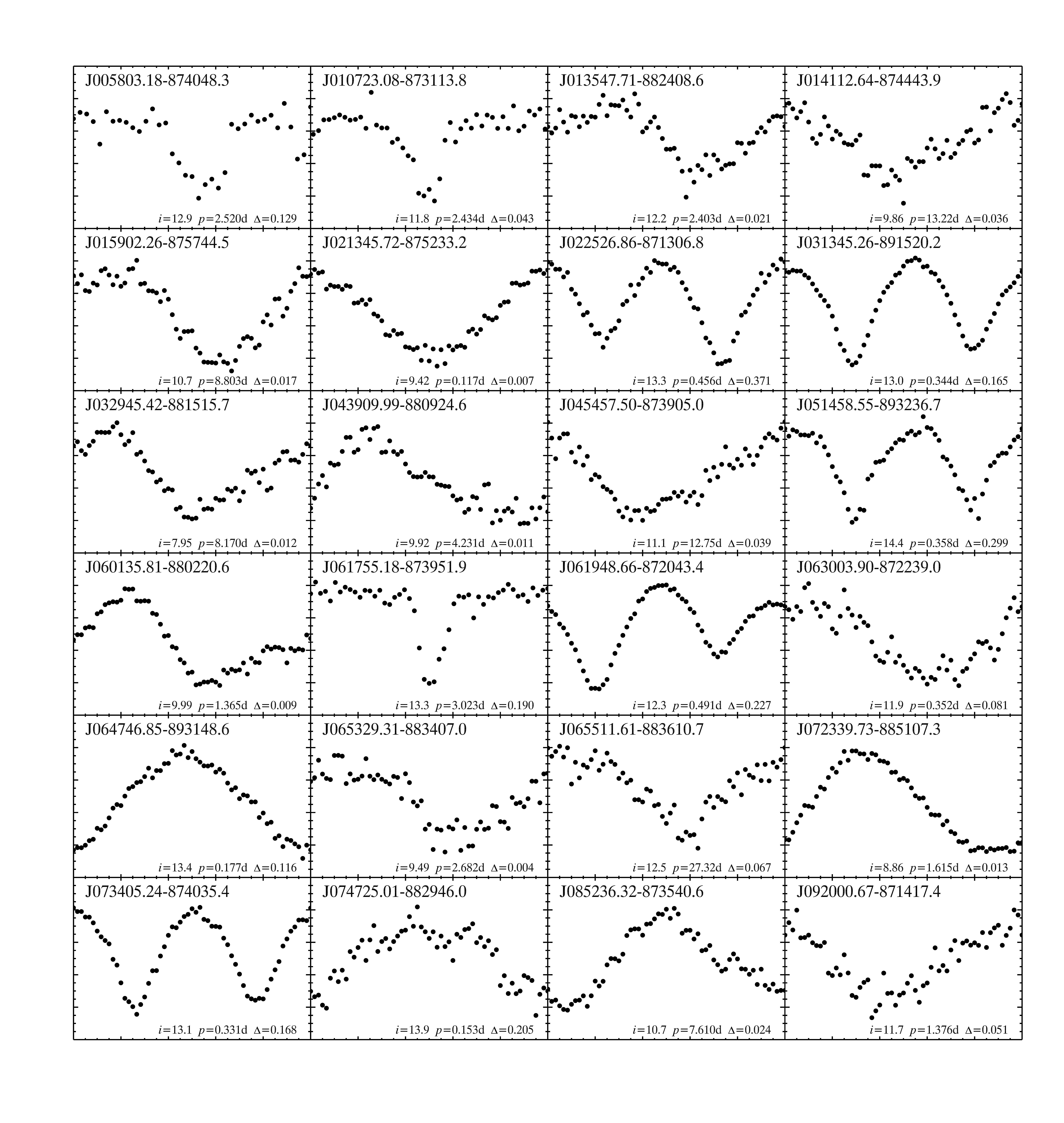}
\caption{$i$-band phased light curves for 60 periodic variables newly discovered in the CSTAR field, in order of increasing right ascension. The identifer for an object together with its median light-curve magnitude, best-determined period, and maximum peak-to-peak amplitude are shown in each panel.
\label{fig9}}
\end{figure}

\addtocounter{figure}{-1}
\begin{figure}
\epsscale{0.75}
\plotone{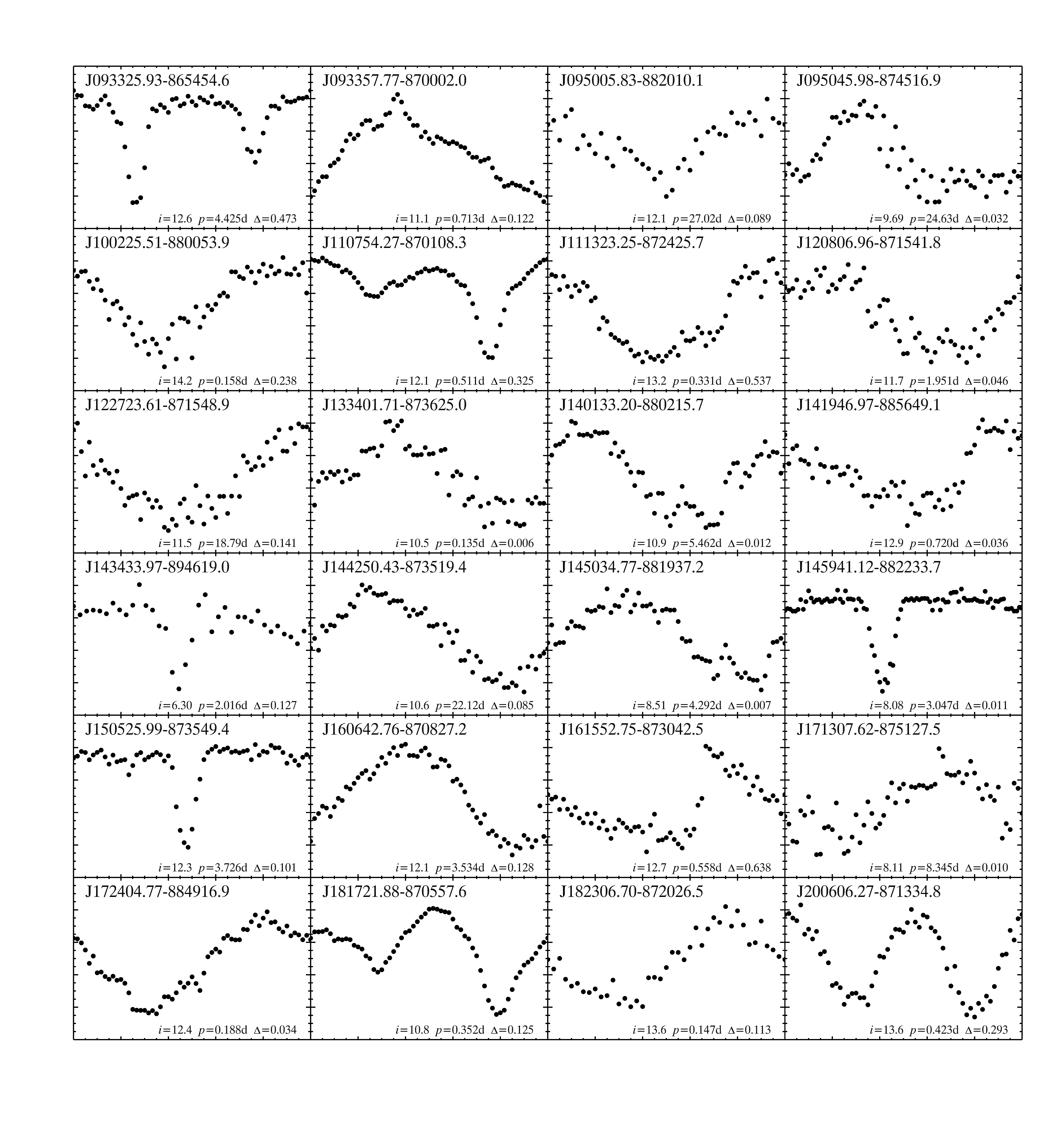}
\caption{Continued
\label{fig9}}
\end{figure}

\addtocounter{figure}{-1}
\begin{figure}
\epsscale{0.75}
\plotone{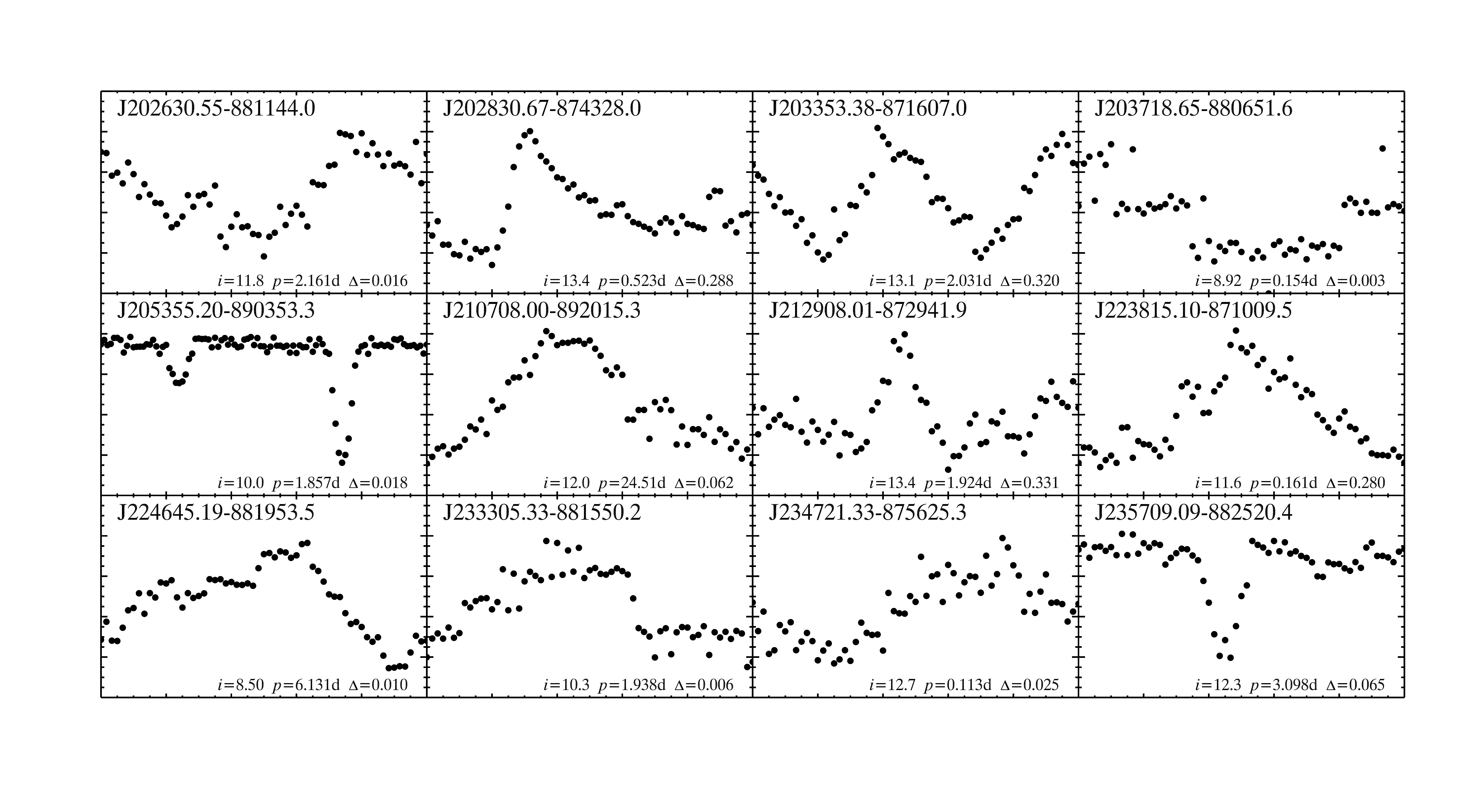}
\caption{Continued
\label{fig9}}
\end{figure}

\clearpage
\begin{figure}
\epsscale{0.75}
\plotone{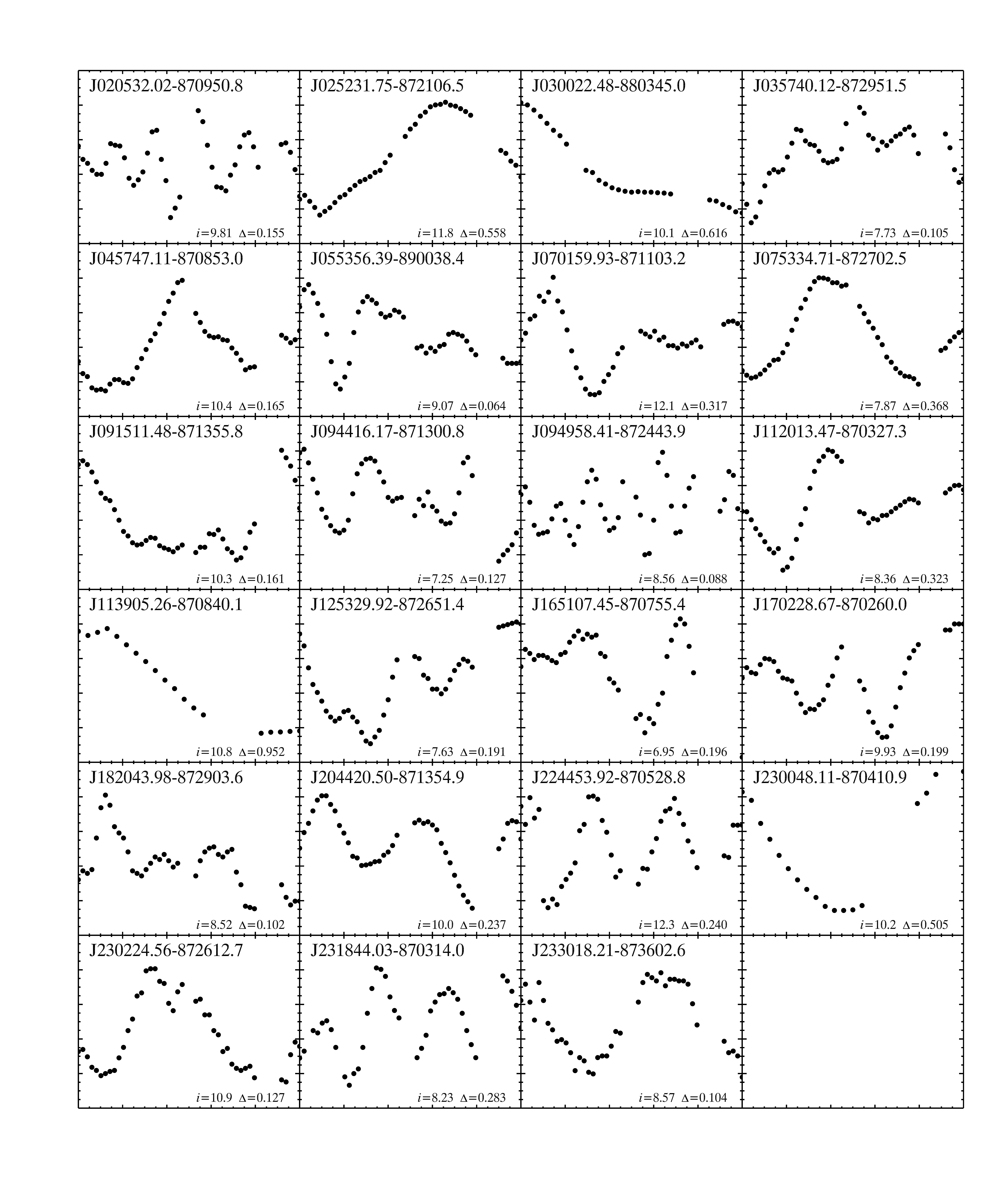}
\caption{$i$-band light curves for 23 non-periodic or quasi-periodic variables newly discovered in the CSTAR field, in order of increasing right ascension. The identifer for an object together with its median light-curve magnitude and maximum peak-to-peak amplitude are shown in each panel.
\label{fig10}}
\end{figure}

\clearpage
\begin{figure}
\epsscale{1.0}
\plotone{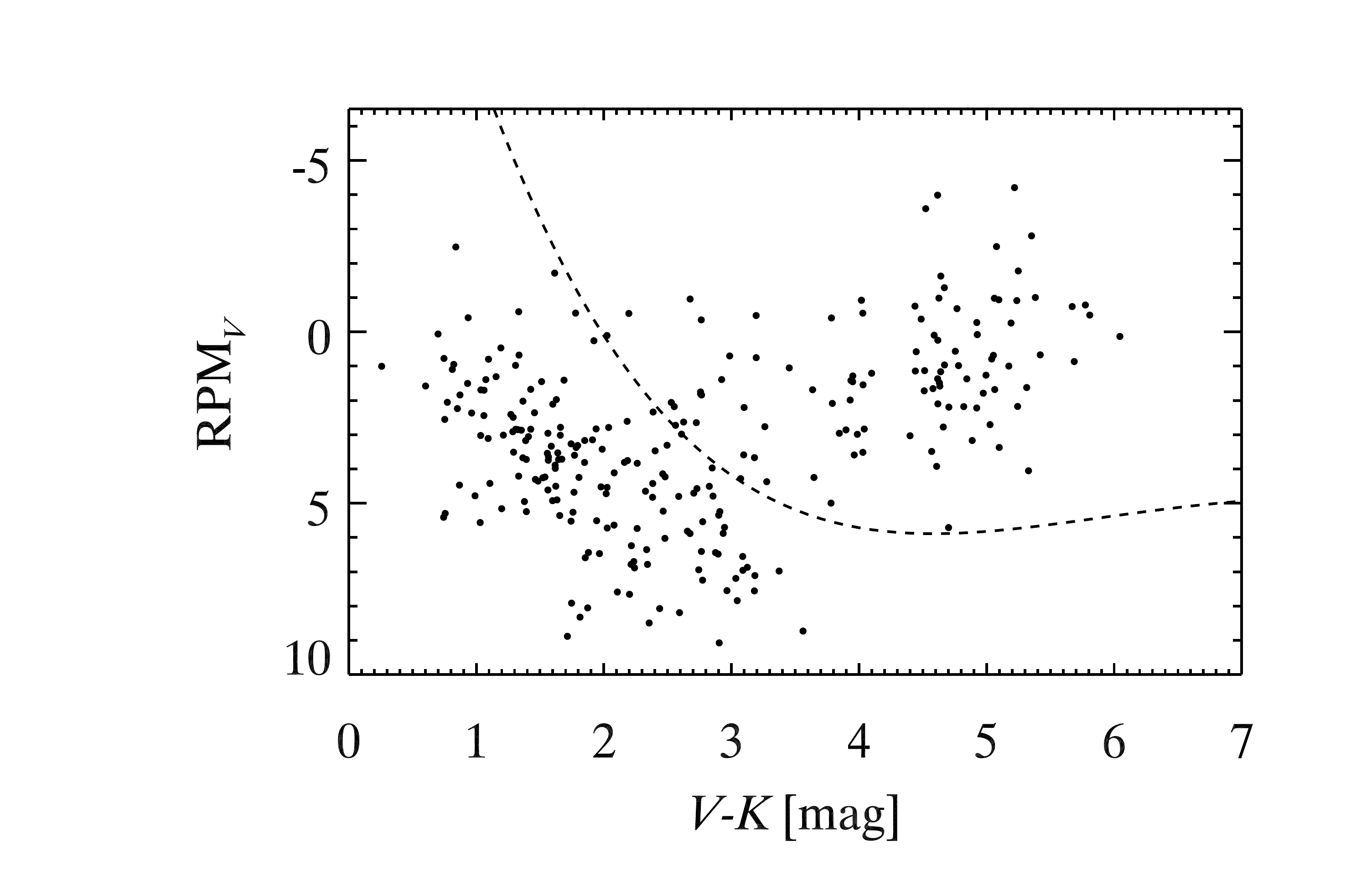}
\caption{$V-K$ vs. reduced proper motion (${\rm RPM}_V$) diagram for the identified variables in the CSTAR field.
Giants are separated from dwarfs, as they lean towards lower of ${\rm RPM}_V$ and higher $V-K$.
A polynomial boundary (the dashed line) separating the two groups is taken from \citet{clarkson2007}.
\label{fig11}}
\end{figure}

\setlength{\tabcolsep}{1.1pt}
\begin{deluxetable}{ccccccccccccccc}
\rotate
\tabletypesize{\scriptsize}
\tablewidth{0pt}
\tablecaption{Catalog of Recovered Variables from the Previous Study in the CSTAR Field}
\tablehead{
\colhead{CSTAR ID}   & \colhead{2MASS ID}  &  \colhead{$i$}      & \colhead{Period}  &  \colhead{Amplitude}  &  \colhead{$T_{\rm 0}$}      &  \colhead{$J$}        &   \colhead{$H$}        &  \colhead{$K$}         &  \colhead{$\mu_{\rm \alpha}$}       &  \colhead{$\mu_{\rm \delta}$}      & \colhead{$T_{\rm eff}$}    & \colhead{Sp.Type}   & \colhead{$ROSAT$ ID}   &  \colhead{Type}\\
\colhead{CSTAR+}      & \colhead{2MASS+}    &  \colhead{(mag)}  & \colhead{(days)}  &  \colhead{(mag)}         &  \colhead{(2454500.0+)}  &\colhead{(mag)}  &  \colhead{(mag)}  & \colhead{(mag)}   & \colhead{(${\rm mas\,yr^{-1}}$ )}      & \colhead{(${\rm mas\,yr^{-1}}$ )}     & \colhead{(K)}        & \colhead{}               & \colhead{1RXS+}       &  \colhead{}\\
}
\startdata
J000032.66-882011.0 &   00005088-8819418 &    9.421 &     0.1144     &     0.010 &    49.2916 &  8.669  & 8.572   &  8.521 & 13.9  & 4.9      &  7134 & A4 III     &  ...&  DSCT*         \\
J000035.85-875505.0 &  00005294-8754270  & 11.622  &     ...             &     0.733 &     ...          &  7.061  & 5.995   & 5.210  &  11.0  & 7.8      & ...       & ...          & ... & IR             \\
J000057.73-874448.1 &  00011717-8744027   &  11.892 &     9.4789     &     0.440 &    57.0879 & 10.983 & 10.66   & 10.594 &   -5.2 & -8.1    & 4664  & G8 IV    & ... & EA             \\
J000826.04-881419.6 &  00084383-8813479   & 10.060 &     ...              &     0.162 &     ...          & 8.000   & 7.106   & 6.814   &   18.2 & 0.8    &...         & ...          & ... & IR             \\
J002022.75-894842.7 &   00202009-8948378   &   9.428 &    10.9234     &     0.025 &    56.8881 & 7.516   & 6.675   & 6.389   &   2.4  &  4.2    & 4763   & M3 III   & ... &   SP*             \\
J002429.26-890902.3 &  00243762-8908470    & 12.600 &     4.2317     &      0.050 &    52.0417 & 12.097  & 11.679 & 11.529 &   45.9 & 69.8   & ...        & ...        & ...   & BY*              \\
J002951.87-893028.7 &  00300000-8930198    & 11.249 &     0.1527     &     0.009  &    49.2987   & 9.557   & 8.799   & 8.581   &   8.2     & 9.8    &  ...      & ...        & ...     & ELL*           \\
J003101.00-885536.1 &  00311591-8855180     & 10.780 &     9.4220     &     0.010 &    53.1070   & 9.887   & 9.595   & 9.533   &  -18.8   & -0.2  & ...       & ...        & ...     &  ELL*             \\
\enddata
\end{deluxetable}

\setlength{\tabcolsep}{1.1pt}
\begin{deluxetable}{ccccccccccccccc}
\rotate
\tabletypesize{\scriptsize}
\tablewidth{0pt}
\tablecaption{Catalog of Newly Discovered Variables in the CSTAR Field}
\tablehead{
\colhead{CSTAR ID}   & \colhead{2MASS ID} &  \colhead{$i$}     & \colhead{Period} &  \colhead{Amplitude}  &  \colhead{$T_{\rm 0}$}        &  \colhead{$J$}        &   \colhead{$H$}       &  \colhead{$K$}        &  \colhead{$\mu_{\rm \alpha}$}       &  \colhead{$\mu_{\rm \delta}$}       & \colhead{$T_{\rm eff}$}   & \colhead{Sp.Type}   & \colhead{$ROSAT$ ID}  &  \colhead{Type}\\
\colhead{CSTAR+}      & \colhead{2MASS+  }  &  \colhead{(mag)} & \colhead{(days)} &  \colhead{(mag)}         &  \colhead{(2454500.0+)}   &   \colhead{(mag)} &  \colhead{(mag)} & \colhead{(mag)}  & \colhead{(${\rm mas\,yr^{-1}}$ )}      & \colhead{(${\rm mas\,yr^{-1}}$ )}     & \colhead{(K)}       & \colhead{}              & \colhead{1RXS+}        &  \colhead{} \\
}
\startdata
J005803.18-874048.3 &  00582819-8739496 & 12.997 &     2.5200 &     0.129 &    52.6134 & 12.299  & 11.830 &  11.708 &   4.8   &  5.7   & ...        & ...         &  ...   &   BY  \\
J010723.08-873113.8 &  01072961-8730248  & 11.806 &     2.4340 &     0.043 &    52.5037 & 10.894  & 10.641 & 10.520 & 16.9   & 14.7   &...        &...          &  ...    &   GD   \\
J013547.71-882408.6 &  01361199-8823525  & 12.204 &     2.4037 &     0.021 &    50.5770 &  11.676  & 11.211  & 11.077 & -75.4 & 31.3   &  ...      &...          &...      &   BY \\
J014112.64-874443.9 &  01412495-8743549  &  9.863 &    13.2200 &     0.036 &    54.5297 &    7.620  &  6.750  &  6.420  &    3.5  &   6.7   &  ...      &...          &...     &   SP  \\
J015902.26-875744.5 &  01590860-8757025  & 10.711 &     8.8030 &     0.017 &    49.3791 &   9.538   &  9.095  &  8.962  &   17.9 & 43.5   &  4603 &  G5 III   &...     & ELL  \\
J020532.02-870950.8 &  02053110-8708532  &  9.818  &       ...       &     0.155 &      ...         &   7.133   &  6.212  &  5.856  &    7.7  &  4.7    &  ...      &...           &...     &   IR \\
J021345.72-875233.2 & 02135122-8751490   &  9.428 &     0.1179 &     0.007 &    49.3093  &   8.583   &  8.399  &  8.388  &   23.9  & -1.7   & 7200  & F0         &...     &   DSCT  \\
J022526.86-871306.8 &  02252557-8712132  & 13.306 &     0.4569 &     0.371 &    50.4089 &  12.525  & 12.264 &12.212  &    2.6   & -0.6   &  ...      &...           &...     &  EW \\
\enddata
\label{table2}
\begin{flushleft}
Notes. Objects are listed in order of increasing right ascension.
       The column descriptions are as follows:
       Col. (1): CSTAR identifier of variability. 
       Col. (2): 2MASS identifier of variability. 
       Col. (3): Median $i$ apparent magnitude.
       Col. (4): Period of variability (only for periodic variables).
       Col. (5): Amplitude of variability.
       Col. (6): Times of minimum brightness of periodic variability (if available).
       Col. (7-9): $JHK$ near-infrared magnitudes from the 2MASS catalog.
       Col. (10, 11): Proper motion in right ascension and declination from the PPMXL catalog.
       Col. (12, 13): Effective temperature and spectral type from the VizieR database (if available). 
       Col. (14): $ROSAT$ identifier of variability. 
       Col. (15): Type of variability: 
                       ACV, $\alpha^2$ Canum Venaticorum variables;
                       BY, BY Draconis-type variables;
                       CEP, Cepheid variables;
                       DCEP, $\delta$ Cephei-type variables;
                       DSCT, $\delta$ Scuti-type variables; 
                       EA, Algol-type eclipsing binaries; 
                       EW, W Ursae Majoris-type eclipsing binaries;
                       ELL, Rotating ellipsoidal variables; 
                       GD, $\gamma$ Doradus variables;
                       IR, Irregular variables;
                       RL, RR Lyrae-type variables; 
                       SP, Spotted variables that are not classified into a particular class;
                       TR, Transit-like events.
\end{flushleft}
\end{deluxetable}

\setlength{\tabcolsep}{0.8pt}
\epsscale{0.8}
\begin{deluxetable}{ccccccccccccccc}
\rotate
\tabletypesize{\scriptsize}
\tablewidth{0pt}
\tablecaption{Missed Variables during Our Variability Search}
\tablehead{
\colhead{CSTAR ID} & \colhead{2MASS ID} & \colhead{$i$}     & \colhead{Period} & \colhead{$\Delta i_{90}$} & \colhead{$T_{\rm 0}$}     & \colhead{$J$}        & \colhead{$H$}        & \colhead{$K$}    &  \colhead{$\mu_{\rm \alpha}$}       &  \colhead{$\mu_{\rm \delta}$} & \colhead{$T_{\rm eff}$}  & \colhead{Sp.Type} & \colhead{$ROSAT$ ID} & \colhead{Type} \\
\colhead{CSTAR+}                  & \colhead{2MASS+}   & \colhead{(mag)} & \colhead{(days)} & \colhead{(mag)}               & \colhead{(2454500.0+)} & \colhead{(mag)} & \colhead{(mag)} & \colhead{(mag)} & \colhead{(${\rm mas\,yr^{-1}}$ )} & \colhead{(${\rm mas\,yr^{-1}}$ )} & \colhead{(K)}    & \colhead{}             & \colhead{1RXS+} & \colhead{}       
}
\startdata
J004526.78-882521.3        & 00452677-8825210 &   11.57  &    34.5266 &     ...        &    ...           &  10.454  &  10.058 &  9.970   &  21.5  &   0.1    &  ...       &  ...      & ... & U          \\
J005607.00-884217.9$^*$ & 00560824-8842175 &   11.36  &      7.9944 &     0.04    &    ...           &  10.278  &    9.773  &  9.672   &  21.7  &   4.4    &  ...       &   ...     & ... & MP         \\
J010804.79-874759.1        & 01080487-8747588 &   13.87  &    15.8462 &     ...        &    ...           &   12.432  &  11.929 &  11.727  &  8.6    &   7.2    &  ...       &   ...     & ... & U          \\
J021102.80-892250.3        & 02110223-8922496  &   10.27  &      6.6230 &     0.03    &    ...           &    8.441  &   7.644  &   7.380   &  9.1     &  31.1  &  ...       & ...       & ...  &MP         \\
J025225.60-872021.5        & 02522563-8720209 &   13.66   &          ...     &     1.03    &    ...           &  5.575    &  3.903   &   2.897   &   -3.6   &  24.9  &  ...      & ...       & ... &IR         \\
J030117.31-892430.2        & 03011730-8924295  &   12.38   &     5.7673  &     0.05    &  785.6845 &  11.321  & 10.930 &  10.842   &   16.4   &  30.6  &       ... & ...       & ... &TR        \\
J031122.58-891513.6        & 03112278-8915130  &   14.08   &     0.1723   &     ...       &     55.6132 &  12.832 & 12.454  & 12.345   & 18.4     & -3.0     & ...      &  ...       & ... &EC         \\
J034410.88-885209.5        & 03441024-8852092 &     11.19  &    21.4962  &     0.03    &    ...           &  9.515   & 8.677    &  8.477    & 7.4       &  3.2      & ...      &...         & ... & MP         \\
\enddata
\label{table3}
\begin{flushleft}
Notes. Objects are listed in order of increasing right ascension.
       The column descriptions are as follows:
       Col. (1): CSTAR identifier of variability, stars marked with asterisks may suffer from 1-d aliases and thus their variability is not confirmed in this study .
       Col. (2): 2MASS identifier of variability. 
       Col. (3): Median $i$ apparent magnitude.
       Col. (4): Period of variability (only for periodic variables). 
       Col. (5): 1.64 standard deviation ($90\%$ confidence interval) of the light curve.
       Col. (6): Times of minimum brightness of periodic variability (if available). 
       Col. (7-9): $JHK$ near-infrared magnitudes from the 2MASS catalog.
       Col. (10, 11): Proper motion in right ascension and declination from the PPMXL catalog.
       Col. (12, 13): Effective temperature and spectral type from the VizieR database (if available). 
       Col. (14): $ROSAT$ identifier of variability. 
       Col. (15): Type of variability: 
                       DSCT, $\delta$ Scuti-type variables; 
                       EC, Contact eclipsing binaries; 
                       ED, Detached eclipsing binaries;  
                       IR, Irregular variables; 
                       MP, Multi-periodic variables; 
                       RL, RR Lyrae-type variables; 
                       TR, Transit-like events; 
                       U, Unclassified variables. 
\end{flushleft}
\end{deluxetable}

\setlength{\tabcolsep}{1.1pt}
\begin{deluxetable}{cccc}
\tabletypesize{\scriptsize}
\tablewidth{0pt}
\tablecaption{Distribution of Variables in the CSTAR Field by Type.}
\tablehead{
\colhead{Type} & \colhead{Recovered Variables} & \colhead{Newly Discovered Variables}     & \colhead{Total Variables}  \\
}
\startdata
ACV          & 2       &   1    &    3     \\
BY             & 35     &   11  &    44     \\
CEP          & 1       &   1    &    2    \\
DCEP        & 2       &   1   &    3     \\
DSCT        & 10     &   7   &    17     \\
EA             & 21     &   6   &    27    \\
EB             & 3       &   0   &    3     \\
EW            & 20     &   8   &    28     \\
ELL           & 13      &  10 &    23     \\
GD            & 12     &   4   &    16     \\
IR              & 30     &  23  &    53     \\
RL             & 6       &   2   &    8     \\
SP             & 34    &   8   &    42    \\
TR             &  2     &   1   &    3     \\
Total          & 191  &   83  &    274     \\
\enddata
\label{table4}
\begin{flushleft}
Notes. The first column is an alphabetical type of variability: 
                       ACV, $\alpha^2$ Canum Venaticorum variables;
                       BY, BY Draconis-type variables;
                       CEP, Cepheid variables;
                       DCEP, $\delta$ Cephei-type variables;
                       DSCT, $\delta$ Scuti-type variables; 
                       EA, Algol-type eclipsing binaries; 
                       EB, $\beta$ Lyrae-type eclipsing binaries; 
                       EW, W Ursae Majoris-type eclipsing binaries;
                       ELL, Rotating ellipsoidal variables; 
                       GD, $\gamma$ Doradus variables;
                       IR, Irregular variables;
                       RL, RR Lyrae-type variables; 
                       SP, Spotted variables that are not classified into a particular class;
                       TR, Transit-like events.         
\end{flushleft}
\end{deluxetable}

\end{document}